\documentclass[conference]{IEEEtran}
\IEEEoverridecommandlockouts
\usepackage{cite}
\usepackage{amsmath,amssymb,amsfonts}
\usepackage{algorithmic}
\usepackage{graphicx}
\usepackage{textcomp}

\usepackage{amsthm}
\usepackage{booktabs} 
\usepackage{multirow} 
\usepackage[table]{xcolor}
\usepackage[colorlinks,linkcolor=red]{hyperref}

\newcommand{\teal}[1]{\textcolor{teal}{#1}}

\def\BibTeX{{\rm B\kern-.05em{\sc i\kern-.025em b}\kern-.08em
    T\kern-.1667em\lower.7ex\hbox{E}\kern-.125emX}}
\begin{document}

\title{Rethinking Medical Anomaly Detection in Brain MRI: An Image Quality Assessment Perspective\\}

\author{
\IEEEauthorblockN{
Zixuan Pan\IEEEauthorrefmark{1},
Jun Xia\IEEEauthorrefmark{1},
Zheyu Yan\IEEEauthorrefmark{1},
Guoyue Xu\IEEEauthorrefmark{1},
Yifan Qin\IEEEauthorrefmark{1},
Xueyang Li\IEEEauthorrefmark{1},
Yawen Wu\IEEEauthorrefmark{1}, 
Zhenge Jia\IEEEauthorrefmark{2}, \\
Jianxu Chen\IEEEauthorrefmark{3},
Yiyu Shi\IEEEauthorrefmark{1}\thanks{Corresponding authors: Jianxu Chen (email: jianxu.chen@isas.de) and Yiyu Shi (email: yshi4@nd.edu).}
}
\IEEEauthorblockA{\IEEEauthorrefmark{1}University of Notre Dame, USA}
\IEEEauthorblockA{\IEEEauthorrefmark{2}Shandong University, China}
\IEEEauthorblockA{\IEEEauthorrefmark{3}Leibniz-Institut für Analytische Wissenschaften – ISAS – e.V., Germany}
}

\maketitle
\begin{abstract} 
Reconstruction-based methods, particularly those leveraging autoencoders, have been widely adopted for anomaly detection task in brain MRI. 
Unlike most existing works try to improve the task accuracy through architectural or algorithmic innovations, we tackle this task from  image quality assessment (IQA) perspective, an under-explored direction in the field.
Due to the limitations of conventional metrics such as  $\ell_1$ in capturing the nuanced differences in reconstructed images for medical anomaly detection, we propose \textit{fusion quality}, a novel metric that wisely integrates the structure-level sensitivity of Structural Similarity Index Measure (SSIM)  with the pixel-level precision of  $\ell_1$.
The metric offers a more comprehensive assessment of reconstruction quality, considering intensity (subtractive property of $\ell_1$ and divisive property of SSIM), contrast, and structural similarity.
Furthermore, the proposed metric makes subtle regional variations more impactful in the final assessment.
Thus, considering the inherent divisive properties of SSIM, we design an \textit{average intensity ratio (AIR)-based data transformation} that amplifies the divisive discrepancies between normal and abnormal regions, thereby enhancing anomaly detection. 
By fusing the aforementioned two components, we devise the IQA approach.
Experimental results on two distinct brain MRI datasets show that our IQA approach significantly enhances medical anomaly detection performance when integrated with state-of-the-art baselines. 
Code is provided \href{https://github.com/zx-pan/MedAnomalyDetection-IQA}{here}.

\end{abstract}


\section{Introduction}
For decades, deep learning methods \cite{DBLP:journals/mia/DongPFYGYSZ22, DBLP:journals/tmi/DongPFXSYSZ23,peng2025u} have been widely used to assist radiologists in disease recognition, such as detecting tumors from brain MRI scans. 
Traditional supervised learning approaches~\cite{sun2022research,zeng2022imagealcapa}, however, require a large amount of labeled data (e.g., tumor segmentation masks), which are often difficult and expensive to obtain for medical images, especially considering the diversity of disease conditions. To address this challenge, many self-supervised, semi-supervised, and weakly supervised learning methods \cite{ pan2024maskeddiffusionselfsupervisedrepresentation} have been developed. These methods effectively utilize both limited labeled data and abundant unlabeled data. Among these approaches, framing the disease recognition task as an anomaly detection problem has gained popularity. This type of method trains solely on unannotated normal images (e.g., MRI scans of healthy brains), enabling the identification of abnormalities (e.g., tumors) without extensive manual annotation.


\begin{figure}[btp]
\centering
\includegraphics[width=\columnwidth]{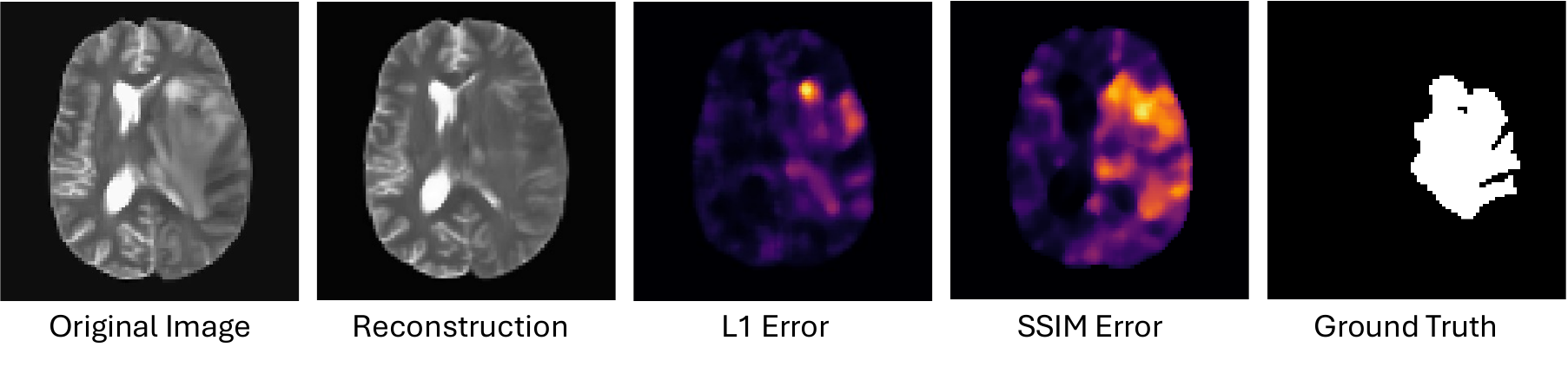} 
\vspace{-0.2in}
\caption{Visualization of the anomaly maps generated by $\ell_{1}$ loss and SSIM loss from the same reconstruction. Calculating the reconstruction discrepancy with L1-metric cannot flag the large tumor area, while calculating with SSIM, from the same reconstruction, could identify the tumor area significantly better.}
\label{ssim}
\vspace{-0.2in}
\end{figure}


Reconstruction-based methods, such as autoencoders (AEs) and their variants, have shown promise in medical anomaly detection. They are trained to reconstruct original images from corrupted inputs, assuming anomalies are harder to reconstruct. During inference, the difference between reconstructed and original images indicates pixel-wise anomaly levels, with abnormal regions exhibiting higher reconstruction errors detectable via post-processing (e.g., thresholding). However, standard AEs and variational autoencoders (VAEs) often produce blurry reconstructions, limiting detection performance. To improve them, methods such as spatial latent dimensions \cite{DBLP:conf/miccai/BaurWAN18}, skip connections \cite{DBLP:conf/isbi/BaurWAN20}, and denoising autoencoders (DAEs) \cite{DBLP:conf/midl/KascenasPO22} have been proposed. Beyond AEs, generative adversarial networks (GANs) \cite{DBLP:journals/mia/SchleglSWLS19,DBLP:conf/ipmi/SchleglSWSL17} and denoising diffusion probabilistic models (DDPMs) \cite{DBLP:conf/midl/BehrendtBKOS23, DBLP:conf/mlmi-ws/IqbalKCH23, DBLP:conf/cvpr/WyattLSW22} have also been applied.

Although these studies focus extensively on architectural and algorithmic improvements, the role of reconstruction evaluation metrics is often overlooked, with most approaches defaulting to $\ell_{1}$ loss. 
In contrast, we revisit the problem of reconstruction-based anomaly detection in brain MRI from the perspective of image quality assessment (IQA), an underexplored aspect in this field.
Our intuition is based on the observation that simply changing how reconstruction residuals are computed can lead to substantial gains in anomaly detection. As shown in Fig.~\ref{ssim}, even with the same reconstruction, compared to the commonly used $\ell_1$ loss, computing anomaly maps with the Structural Similarity Index Measure (SSIM), a widely adopted IQA metric, can uncover subtle anomalies that would otherwise be overlooked.

Based on the above observations, we argue that metrics beyond $\ell_{1}$ are essential for a more comprehensive assessment of reconstructions during the training and inference phases of anomaly detection.
Therefore, we propose a novel image quality-based assessment metric named \textit{fusion quality}   that wisely combines both SSIM (structure-level quality) and the widely used $\ell_{1}$ (pixel-level quality). 
This combined metric evaluates the reconstruction based on intensity (subtractive from $\ell_1$ and divisive from SSIM), contrast, and structure similarity, adaptively capturing the strength of both quality assessment metrics.

Evaluating reconstruction quality beyond just pixel-wise intensity introduces a higher level view, making subtle variations in different regions more impactful in the final assessment, compared to existing anomaly detection solutions. In this situation, the inherent characteristics of images from a semantic perspective become increasingly important for optimizing anomaly detection performance, therefore necessitates commensurate pre-processing steps tailored to these expanded metrics. 
Since SSIM in the proposed fusion quality measurement is designed in a divisive way (see Equation~\eqref{ssim_lcs}), it is important to amplify the divisive discrepancies between anomalies and normal regions.
To this end, we propose an \textit{average intensity
ratio-based data transformation} to consistently enhance the divisive discrepancies between normal and abnormal regions, thereby improving the overall effectiveness of the model.

We refer to our final approach, which combines the fusion quality loss and AIR enhancement pre-processing strategy, as the IQA approach. We evaluate its effectiveness on several commonly used datasets by applying it to a baseline model.

We summarize our main contributions as follows:
\begin{itemize}
    
    \item \textbf{IQA-inspired Loss and AIR-based Transformation}: To the best of our knowledge, we are the \textit{first} to use a comprehensive evaluation metric, \textit{fusion quality loss}, which incorporates SSIM loss alongside $\ell_{1}$ loss for both training and inference in brain MRI anomaly detection. We also propose a simple yet effective \textit{average
intensity ratio-based data transformation} to enhance the divisive discrepancie between normal and abnormal regions, and validate its effectiveness empirically.

    \item \textbf{Strong Empirical Results}: Our results show that our method achieves relative improvements in DICE of up to 15.86\% for BraTS21 T2, 21.41\% for MSLUB T2 compared to state-of-the-art (SOTA) baselines. We also show that the proposed method can well generalize to other modalities and backbones.

    \item\textbf{Image Quality Assessment (IQA) Perspective}: We investigate brain MRI anomaly detection from an image quality assessment perspective and achieve state-of-the-art performance on the BraTS21 and MSLUB datasets. Our approach opens a new door in the community for studying medical image anomaly detection.
\end{itemize}

\section{Related Work}
In recent years, reconstruction-based methods using autoencoders (AEs) and their variants have become popular for medical anomaly detection, as they model normal anatomy without requiring abnormal labels. These models reconstruct healthy images, using reconstruction error as an anomaly score. However, AEs and VAEs often produce blurry reconstructions, limiting anomaly detection \cite{DBLP:journals/mia/BaurDWNA21}. To address this, advanced AE models have been proposed: vector-quantized VAEs \cite{MIA/PINAYA} improve discrete feature representation, adversarial autoencoders \cite{Chen2018UnsupervisedDO} enhance generative quality via adversarial training, and denoising autoencoders (DAE) \cite{DBLP:conf/midl/KascenasPO22} improve image clarity with skip connections and denoising tasks.

Other than AE-based methods, generative adversarial networks (GANs) have also been applied. 
AnoGAN \cite{DBLP:conf/ipmi/SchleglSWSL17}, the first GAN-based approach for this task, detects anomalies by comparing test images to GAN-generated healthy counterparts. However, AnoGAN requires extensive inference time due to its reliance on numerous back-propagation iterations. To improve inference speed, f-AnoGAN \cite{DBLP:journals/mia/SchleglSWLS19} uses an encoder with a Wasserstein GAN for faster mapping to latent space. Despite these improvements, GANs still still face stability and anatomical coherence issues \cite{DBLP:journals/mia/BaurDWNA21}.

Denoising diffusion probabilistic models (DDPM) have recently gained attention as a robust method for anomaly detection in brain MRI. anoDDPM \cite{DBLP:conf/cvpr/WyattLSW22} was the first to apply DDPM in this context, proposing the use of simplex noise to replace Gaussian noise. Building on this, pDDPM \cite{DBLP:conf/midl/BehrendtBKOS23} improved anomaly detection performance by adopting a patch-based DDPM approach, where noise is added to patches while the rest of the image remains uncorrupted and serves as a condition. This technique enhances brain MRI reconstruction by incorporating global context information about individual brain structures and appearances. Further extending this concept, mDDPM \cite{DBLP:conf/mlmi-ws/IqbalKCH23} applied the patch-based approach to the frequency domain, yielding additional improvements.

While much of work in the anomaly detection has focused on designing architectures and algorithms, some studies have investigated different ways of measuring discrepancies. For instance, \cite{DBLP:conf/visapp/BergmannLFSS19} applies SSIM loss for industrial defect detection, and replacing the $\ell_2$ loss. \cite{DBLP:conf/brainles-ws/MeissenPKR22} proposed calculating SSIM loss in latent space instead of pixel space, and \cite{behrendt2024diffusionmodelsensembledstructurebased} 
designed an ensembled SSIM approach for anomaly score calculation.

In summary, prior work either applies SSIM in latent space or only at inference.
 We are the first to use a comprehensive evaluation metric that incorporates SSIM loss alongside $\ell_{1}$ loss for both training and inference in medical anomaly detection problem, achieving state-of-the-art performance on several commonly used datasets.

\section{Method}
\begin{figure}[btp]
\centering
\includegraphics[width=1\columnwidth]{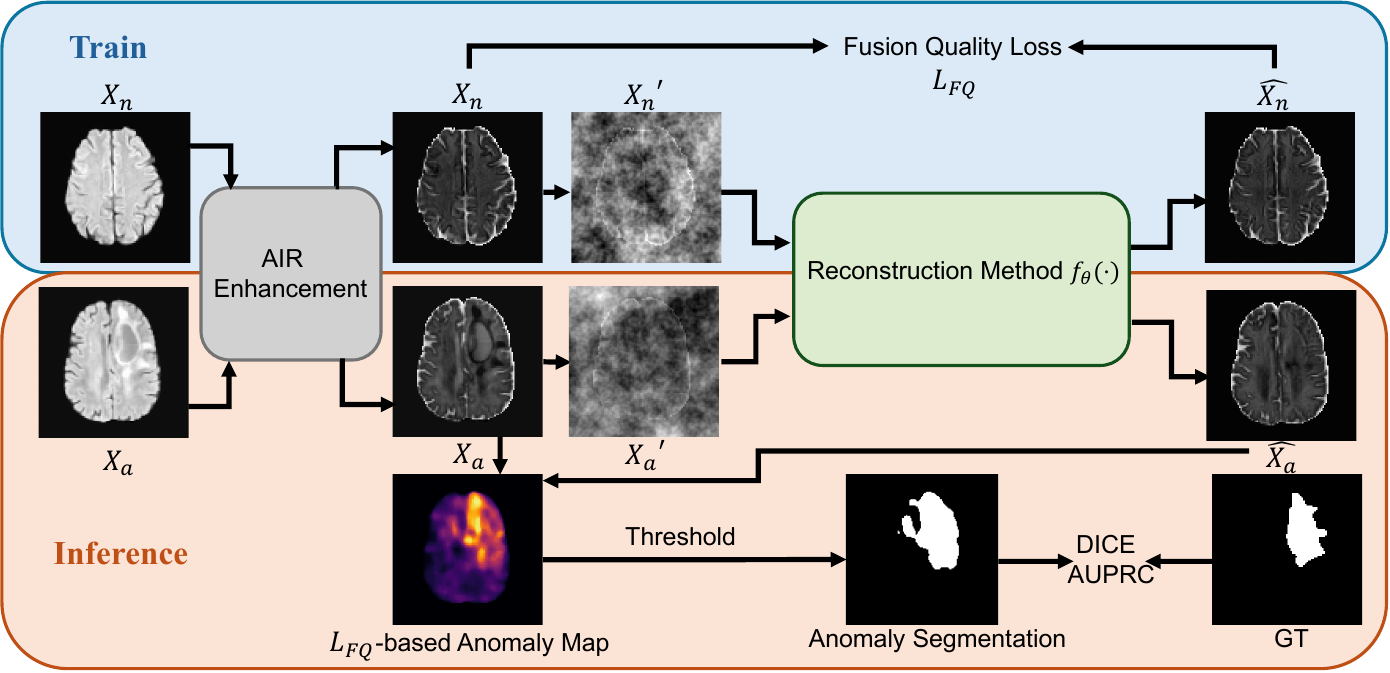} 
\caption{Overview of our reconstruction-based anomaly detection method with the proposed \textit{fusion quality loss} and \textit{AIR-based data transformation}. 
During training, the normal dataset $X_n$ is augmented with the proposed AIR-based data transformation to enhance the divisive discrepancies, and corrupted to form the noisy normal dataset $X_{n}{'}$ using simplex noise.
The model is then trained by denoising $X_{n}'$ and minimizing the fusion quality loss $L_{FQ}$ between the reconstruction $\hat{X_n}$ and the original normal dataset $X_n$. During inference, the abnormal test dataset $X_a$ undergoes the same process.
The anomalies in $X_a$ are expected to be poorly reconstructed, resulting in higher values in the $L_{FQ}$-based anomaly map.
The final anomaly map is thresholded for segmentation, with performance measured in terms of DICE and AUPRC.
}

\label{model}
\vspace{-0.1in}
\end{figure}

In this section, we will first sketch the overall framework of reconstruction-based anomaly detection. We then introduce our proposed \textit{Fusion Quality Loss} and \textit{Average Intensity Ratio-based Transformation}, two major findings after revisiting the brain MRI anomaly detection from an image quality assessment (IQA) perspective. An overview of our final reconstruction-based anoamly detection framework is shown in Fig.~\ref{model}.

\subsection{Reconstruction-based Anomaly Detection}
\label{uad}
Let  $\mathbf{X}^n = \{ \mathbf{x}_i^n \in \mathcal{X}^n\}_{i=1}^N$  represent $N$ samples in a normal data space $\mathcal{X}^n$. Reconstruction-based anomaly detection aims to train a model  $f_\theta(\cdot)$  that reconstructs $\mathbf{x}_i^n$ from a corrupted version $\mathbf{x}_i^{n\prime}$ by minimizing a reconstruction loss:
\vspace{-0.1in}
\begin{equation}
    \min_{\theta} \frac{1}{N} \sum_{i=1}^N L_{\text{train}}\big(\mathbf{x}_i^n, \hat{\mathbf{x}}_i^n\big), \quad \text{where } \hat{\mathbf{x}}_i^n = f_\theta\big(\mathbf{x}_i^{n\prime}\big).
\end{equation}
$L_{\text{train}}$ is a function to measure reconstruction quality.
During test, for a test image $\mathbf{x}^{a}_{j} \in \mathbf{X}^{a} = \{ \mathbf{x}_j^a \in \mathcal{X}^a \}_{j=1}^M$, we first degrade it to $\mathbf{x}^{a}_{j}{'}$, and then use the trained reconstruction model $f_{\theta^{*}}(\cdot)$ to get the reconstruction $\hat{\mathbf{x}^{a}_{j}}$.
The pixel-wise anomaly score map $\Lambda_{j}$ is defined by the reconstruction error:
\begin{equation}
\Lambda_j = L_{\text{test}}\big(\mathbf{x}_j^a, \hat{\mathbf{x}}_j^a\big), \quad \text{where } \hat{\mathbf{x}}_j^a = f_\theta^*\big(\mathbf{x}_j^{a\prime}\big).
\end{equation}
Higher values in $\Lambda_{j}$ correspond to larger reconstruction errors, indicating a higher probability of abnormality.
$L_\text{test}$ serves the same purpose of assessing the reconstructions as $L_\text{train}$, though it may use a different function. A threshold is then applied to $\Lambda_{j}$ for binarization, yielding the final anomaly segmentation.

\subsection{Fusion Quality Loss}
Most existing reconstruction-based anomaly detection methods in Brain MRI use $\ell_{1}$ loss to calculate the reconstruction error during training and test.
However, $\ell_{1}$ loss has two main issues in anomaly detection problems:
it assumes pixel independence, ignoring spatial relationships, which may prevent the model from learning the intrinsic structure of healthy brains.
Additionally, it focuses on intensity discrepancies, which may not capture subtle anomalies with only minor intensity differences from normal parts.

To address these limitations, we propose to assess the reconstruction quality from a more comprehensive perspective by incorporating the Structural Similarity Index Measure (SSIM), a widely used and differentiable metric in image quality assessment (IQA). While other perceptual metrics such as LPIPS are also popular in natural image tasks, they typically require deep pretrained networks, which may not generalize well to medical images. In contrast, SSIM captures luminance, contrast, and structural differences in a lightweight and interpretable manner, making it more suitable for our setting. Moreover, its differentiability ensures compatibility with gradient-based training.

SSIM is originally constructed as an image quality measure reflecting human perception rather than absolute differences like Mean Squared Error (MSE). It assesses similarity between two images $\mathbf{x}$ and $\mathbf{y}$ across three components: luminance $l(\mathbf{x}, \mathbf{y})$, contrast $c(\mathbf{x}, \mathbf{y})$, and structure $s(\mathbf{x}, \mathbf{y})$, defined as:
\begin{equation}
\begin{split}
l(\mathbf{x}, \mathbf{y}) &= \frac{2\mu_\mathbf{x}\mu_\mathbf{y} + C_1}{\mu_\mathbf{x}^2 + \mu_\mathbf{y}^2 + C_1}, \quad
c(\mathbf{x}, \mathbf{y}) = \frac{2\sigma_\mathbf{x}\sigma_\mathbf{y} + C_2}{\sigma_\mathbf{x}^2 + \sigma_\mathbf{y}^2 + C_2},\\
s(\mathbf{x}, \mathbf{y}) &= \frac{\sigma_{\mathbf{xy}} + C_3}{\sigma_\mathbf{x}\sigma_\mathbf{y} + C_3},
\end{split}
\label{ssim_lcs}
\end{equation}
where \( \mu_\mathbf{x} \) and \( \mu_\mathbf{y} \) are the means of the images \( \mathbf{x} \) and \( \mathbf{y} \), respectively. \( \sigma_\mathbf{x} \) and \( \sigma_\mathbf{y} \) are the standard deviations of \( \mathbf{x} \) and \( \mathbf{y} \), respectively. \( \sigma_{xy} \) is the covariance between \( \mathbf{x} \) and \( \mathbf{y} \). \( C_1, C_2, \) and \( C_3 \) are small constants for numerical stability.
SSIM is then computed as the product of these three components:
\vspace{-0.05in}
\begin{equation}
\begin{aligned}
    \text{SSIM}(\mathbf{x}, \mathbf{y}) &= l(\mathbf{x}, \mathbf{y}) \cdot c(\mathbf{x}, \mathbf{y}) \cdot s(\mathbf{x}, \mathbf{y}) \\
    &= \frac{(2\mu_\mathbf{x}\mu_\mathbf{y}+C_1)(2\sigma_{\mathbf{xy}}+C_2)}{(\mu_{\mathbf{x}}^{2} + \mu_{\mathbf{y}}^{2} + C_1)(\sigma_{\mathbf{x}}^{2} + \sigma_{\mathbf{y}}^{2} + C_2)}. 
\end{aligned}
\end{equation}

In practice, it is useful to apply SSIM index locally rather than globally for many reasons. The most straightforward one for anomaly detection is that we need a spatially varying quality map of the reconstruction image to localize the anomalies. The local statistics $\mu_\mathbf{x}$, $\sigma_\mathbf{x}$ and $\sigma_\mathbf{xy}$ are calculated within a $W \times W$ window, moving with a stride $S$ over the entire image. We set $W = 5$ and $S = 1$ to produce a quality map matching the input shape. The final SSIM loss between an image $\mathbf{x}$ and its reconstruction $\hat{\mathbf{x}}$ is defined as:
\begin{equation}
    L_\text{SSIM}(\mathbf{x}, \hat{\mathbf{x}}) = \frac{1 - \frac{1}{K}\sum_{k=1}^{K}\text{SSIM}(\mathbf{x}_k, \hat{\mathbf{x}}_k)}{2},
\end{equation}
where $\mathbf{x}_k$ and $\hat{\mathbf{x}}_k$ are the image patches in the $k$-th local window, and $K$ is the total number of windows.
The error at the \((i, j)\) pixel during inference is defined as:
\begin{equation}
    \Lambda(i, j) = \frac{1 - \text{SSIM}(\mathbf{x}_{ij}, \hat{\mathbf{x}}_{ij})}{2},
\end{equation}
where \(\mathbf{x}_{ij}\) and \(\hat{\mathbf{x}}_{ij}\) are local image patches centered at \((i, j)\).

By design, SSIM is not particularly sensitive to uniform biases, which can lead to changes in brightness or color shifts. However, SSIM better preserves contrast in high-frequency regions compared to other loss functions as shown in \cite{DBLP:journals/tci/ZhaoGFK17}. Conversely, $\ell_1$ loss maintains color and luminance consistency but lacks structural awareness and contrast preservation.

Recognizing the complementary nature of the two error functions, we design a novel Fusion Quality Loss which wisely combines their strengths:
\vspace{-0.1in}
\begin{equation}
    L_\text{FQ} = \alpha L_\text{SSIM} + (1 - \alpha) L_{\ell_1}, \quad \alpha \in [0, 1].
\end{equation}
We set $\alpha = 0.84$ without further tuning, as suggested by prior work \cite{DBLP:journals/tci/ZhaoGFK17}. More discussions are in Section~\ref{alpha_albation}.

\subsection{Average Intensity Ratio-based Transformation}
After incorporating SSIM loss into the reconstruction assessment metric, the error is no longer uniformly weighted regardless of the local structure as it is with $\ell_1$ loss. Instead, the structural relationships between regions become more significant, making anomaly detection more sensitive to dataset characteristics.
Moreover, since the proposed fusion quality loss introduces divisive components from SSIM, amplifying the divisive discrepancies between anomalies and normal regions becomes crucial.
To further enhance anomaly detection performance under this new loss function, we propose an image processing transformation called \textit{average intensity ratio (AIR)-based  transformation} that optimally reinforces these divisive discrepancies.
We define the average intensity ratio (AIR) between the anomalous and normal regions in an abnormal dataset $\mathbf{X}$ as:
\begin{equation}
\label{AIR}
\mathrm{AIR}(\mathbf{X}) = \frac{(\mu_{\mathbf{X}}^a + \mu_{\mathbf{X}}^n) + |\mu_{\mathbf{X}}^a - \mu_{\mathbf{X}}^n|}{(\mu_{\mathbf{X}}^a + \mu_{\mathbf{X}}^n) - |\mu_{\mathbf{X}}^a - \mu_{\mathbf{X}}^n|},
\end{equation}
where $\mu^{a}_{\mathbf{X}}$ and $\mu^{n}_{\mathbf{X}}$ are the mean pixel intensities of the anomalous and normal regions, respectively, defined as:
\begin{equation}
\mu_{\mathbf{X}}^t = \frac{1}{N} \sum_{k=1}^N \frac{1}{|\mathcal{R}^k|} \sum_{(i, j) \in \mathcal{R}^k} I(x_{ij}^k) P^t(x_{ij}^k), \quad t \in {a, n}
\end{equation}
where $t = a$ for anomalous regions and $t = n$ for normal regions, $N$ is the total number of images in $\mathbf{X}$, $\mathcal{R}^k$ is the pixel set in the $k$-th image, $I(x_{ij}^k)$ represents the intensity of pixel $(i, j)$, and $P^t(x_{ij}^k)$ is the probability measure indicating whether the pixel belongs to a normal ($t=n$) or anomalous ($t=a$) region.

Based on the principles of reconstruction-based anomaly detection sketeched in Section~\ref{uad}, our transformation aims to increase AIR of the dataset, as a higher AIR indicates greater discrepancies between normal training data and test anomalies, resulting in larger generalization errors in the abnormal regions.
Existing baselines use a small validation set $\mathbf{X}_\text{val} \subset \mathbf{X}^{a}$ and its ground truth $\mathbf{Y}_\text{val}$ for hyperparameter selection (e.g., binarization threshold).
Thus, it is feasible to perform dataset statistics-based pre-processing transformation before training to increase AIR and improve anomaly detection.

In the context of MRI brain anomaly detection, we analyze four modalities of the BraTS dataset, and propose a simple yet effective way that consistently increases AIR. Based on validation set statistics: 1) $0 < \mu_{\mathbf{X}}^{n} < \mu_{\mathbf{X}}^{a} < 1$ across all four modalities; 2) $\mu_{\mathbf{X}}^{n} > 0.5$ for T1, FLAIR and T1-CE; 3) $\mu_\mathbf{X}^a < 0.5$ for T2,
we define AIR-based transformation $p$ as:
\begin{equation}
    p(\mathbf{x}) = \mathbf{x} \cdot \mathbb{I}(\mu_\mathbf{X}^n \leq 0.5) + (\mathbf{1} - \mathbf{x}) \cdot \mathbb{I}(0.5 < \mu_\mathbf{X}^n),
\end{equation}
where $\mathbb{I}$ is an indicator function that returns 1 if the condition inside is true and 0 otherwise. 
Note that in our experiments, the processing is applied only to the non-zero foreground. We omit this detail here for simplicity in writing.
It can be formally proven that this transformation ensures $\mathrm{AIR}(\mathbf{\bar{X}}) \geq \mathrm{AIR}(\mathbf{X})$ for the transformed dataset $\mathbf{\bar{X}}$.

Finally, we refer to our approach as the IQA approach, including the proposed \textit{Fusion Quality Loss} and \textit{Average Intensity Ratio-based Transformation} as its two key components.

\begin{figure}[btp]
\centering
\includegraphics[width=1\columnwidth]{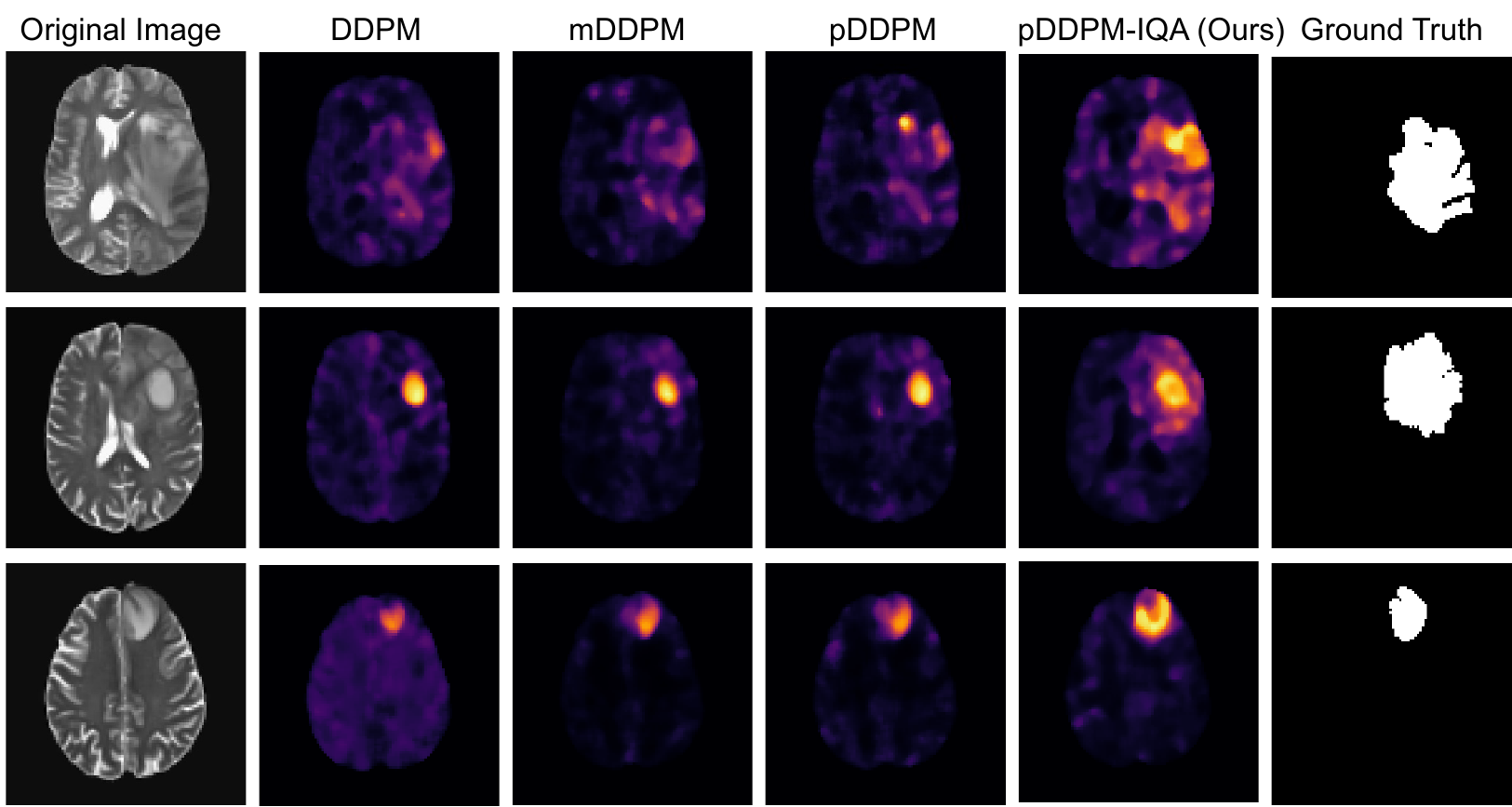} 
\caption{Qualitative visualization on the BraTS21 test set. Columns 2-5 show anomaly maps from different methods for three samples.}
\label{results_t2_fig}
\vspace{-0.1in}
\end{figure}

\begin{table}[!t]
    \centering
    \caption{Comparison with baselines in terms of DICE and AUPRC on BraTS and MSLUB using T2 modality in a \textbf{cross-dataset} setting. The model is trained on the IXI dataset containing only healthy samples. Best results for a given metric/dataset are \textbf{bolded}, while second-best ones are \underline{underlined}.}
    \resizebox{\linewidth}{!}{
    \begin{tabular}{lcccc}
        \toprule
         \multirow{2}{*}{Method} &\multicolumn{2}{c}{BraTS21 (T2)} &\multicolumn{2}{c}{MSLUB (T2)}\\
         \cmidrule(lr){2-3} \cmidrule(lr){4-5}
        & DICE [\%] & AUPRC [\%] &DICE [\%] &AUPRC [\%]\\
        \midrule
        Thresh \cite{DBLP:conf/brainles-ws/MeissenKR21} &19.69 &20.27 &6.21 &4.23\\
        \midrule
        AE \cite{DBLP:journals/mia/BaurDWNA21} &32.87$\pm$1.25 & 31.07$\pm$1.75 &7.10$\pm$0.68 &5.58$\pm$0.26\\
        VAE \cite{DBLP:journals/mia/BaurDWNA21} &31.11$\pm$1.50 &28.80$\pm$1.92 &6.89$\pm$0.09 &5.00$\pm$0.40\\
        SVAE \cite{behrendt2022capturing} &33.32$\pm$0.14 &33.14$\pm$0.20 &5.76$\pm$0.44 &5.04$\pm$0.13\\
        DAE \cite{DBLP:conf/midl/KascenasPO22} &37.05$\pm$1.42 &44.99$\pm$1.72 &3.56$\pm$0.91 &5.35$\pm$0.45\\
        f-AnoGAN \cite{DBLP:journals/mia/SchleglSWLS19} &24.16$\pm$2.94 &22.05$\pm$3.05 &4.18$\pm$1.18 &4.01$\pm$0.90\\
        DDPM \cite{DBLP:conf/cvpr/WyattLSW22} &40.67$\pm$1.21 &49.78$\pm$1.02 &6.42$\pm$1.60 &7.44$\pm$0.52\\
        mDDPM \cite{DBLP:conf/mlmi-ws/IqbalKCH23} &\underline{51.31$\pm$0.66} &\underline{57.09$\pm$0.94} &8.08$\pm$0.70 &9.06$\pm$0.62\\
        pDDPM \cite{DBLP:conf/midl/BehrendtBKOS23} &49.41$\pm$0.66 &54.76$\pm$0.83 &\underline{10.65$\pm$1.05} &\underline{10.37$\pm$0.51}\\
        \midrule
pDDPM-IQA (ours) &\textbf{59.45$\pm$0.37}  &\textbf{62.99$\pm$0.37}
&\textbf{12.93$\pm$0.67} &\textbf{11.51$\pm$0.50}\\ 
$\Delta$ (Relative improvements)   & \teal{(20.32↑)} & \teal{(15.03↑)} & \teal{(21.41↑)} & \teal{(10.99↑)} \\
        \bottomrule
    \end{tabular}
    }
    \label{results_t2}
    \vspace{-0.15in}
\end{table}

\begin{figure*}[!t]
\centering
\includegraphics[width=1\textwidth]{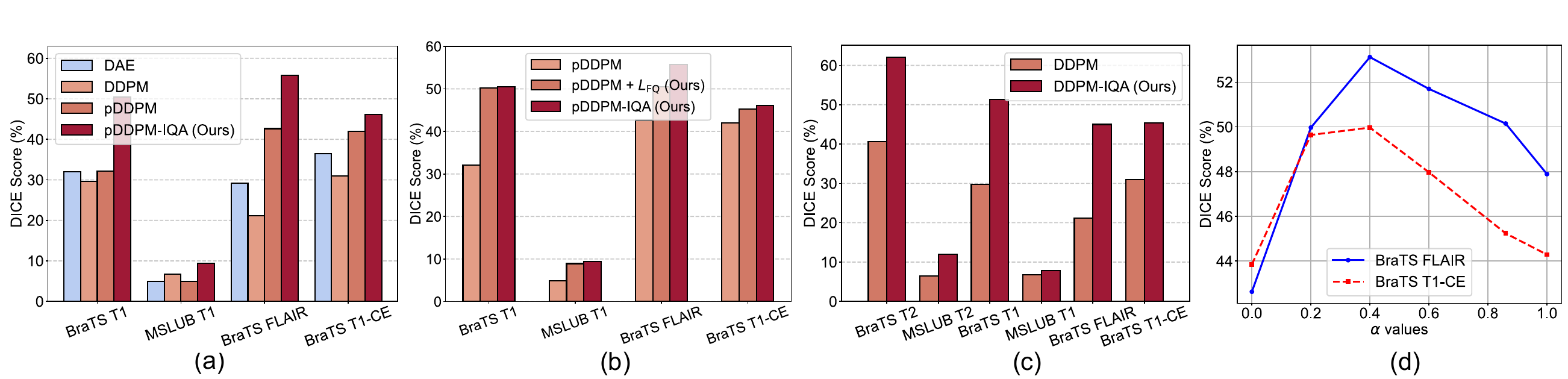} 
\caption{Ablation Study Results.}
\label{ablation_fig}
\end{figure*}

\section{Experiments}
\subsection{Datasets and Implementation Details}
We conduct experiments under both cross-dataset and intra-dataset settings using three public datasets: the Multimodal Brain Tumor Segmentation Challenge 2021 (BraTS21) \cite{baid2021rsna}, the multiple sclerosis dataset from University Hospital of Ljubljana (MSLUB) \cite{DBLP:journals/ni/LesjakGKLPLS18}, and the IXI dataset \cite{ixi}. BraTS21 contains 1251 brain MRI scans with four modalities (T1, T1-CE, T2, FLAIR). MSLUB consists of scans from 30 multiple sclerosis (MS) patients with T1, T2, and FLAIR-weighted images. IXI includes 560 T1–T2 scan pairs of healthy brains.

In the \textbf{cross-dataset} setting, following \cite{DBLP:conf/midl/BehrendtBKOS23}, we perform five-fold cross-validation, training on healthy T1/T2-weighted scans from IXI and evaluating on T1/T2 scans from BraTS21 and MSLUB. 

In the \textbf{intra-dataset} setting, five-fold cross-validation is performed on FLAIR and T1-CE scans from BraTS21. For each fold, slices without tumors from 60\% and 10\% of patients are used for training and training-phase validation, respectively; the remaining 30\% are split into 10\% unhealthy validation and 20\% test sets. All datasets are pre-processed with resampling, skull-stripping, and registration, following \cite{DBLP:conf/midl/BehrendtBKOS23}.

We train the models on NVIDIA A10 GPUs using the Adam optimizer, with a learning rate of 1e-4 and a batch size of 32. We use the default settings in pDDPM \cite{DBLP:conf/midl/BehrendtBKOS23} including using simplex noise as suggested in \cite{DBLP:conf/cvpr/WyattLSW22}, uniformly sampling noise levels $t \in [1, T]$ with $T = 1000$ during training, and training for 1600 epochs. For evaluation, we set the noise level $t_{test}$ to 500 for BraTS21 (T2) and 750 for the others. 
To refine anomaly
maps, we employ standard post-processing for \textit{all} methods, ensuring
optimal performance for each method. First, we apply a median filter with a kernel size of \(K_M = 5\) to smooth anomaly scores, followed by three iterations of brain mask erosion. To determine the optimal binarization threshold, we perform a greedy search on the unhealthy validation set, iteratively calculating Dice scores for various thresholds. The best threshold identified is then used to compute Dice and AUPRC on the unhealthy test set.

\subsection{Comparisons with State-of-the-art Methods}
We apply our IQA approach to a strong baseline pDDPM and compare it against Thresh \cite{DBLP:conf/brainles-ws/MeissenKR21}, AE \cite{DBLP:journals/mia/BaurDWNA21}, VAE \cite{DBLP:journals/mia/BaurDWNA21}, SVAE \cite{behrendt2022capturing}, DAE \cite{DBLP:conf/midl/KascenasPO22}, f-AnoGAN \cite{DBLP:journals/mia/SchleglSWLS19}, DDPM \cite{DBLP:conf/cvpr/WyattLSW22}, mDDPM \cite{DBLP:conf/mlmi-ws/IqbalKCH23} and pDDPM \cite{DBLP:conf/midl/BehrendtBKOS23}, in terms of Dice-Coefficient (DICE) and the average Area Under the Precision-Recall Curve (AUPRC). Results are reported as ``mean$\pm$std'' across five folds.

In Table~\ref{results_t2}, we compare our pDDPM-IQA with state-of-the-art methods on BraTS21 and MSLUB using T2 modality in a \textbf{cross-dataset} setting, as adopted in previous studies \cite{DBLP:conf/midl/BehrendtBKOS23, DBLP:conf/mlmi-ws/IqbalKCH23}. Our pDDPM-IQA significantly (p $<$ 0.05) outperforms all baseline approaches on both datasets in terms of DICE and AUPRC, with improvements exceeding 10\%.
Qualitative examples of anomaly maps generated by our method and other models are shown in Fig.~\ref{results_t2_fig}, demonstrating that pDDPM-IQA provides more precise anomaly detection.

\subsection{Ablation Study}

\noindent \textbf{Performance across Multiple Modalities.} As shown in Fig.~\ref{ablation_fig} (a), we systematically evaluate our method
on a range of MRI modalities, 
including BraTS T1 and MSLUB T1 in a cross-dataset setting, as well as BraTS FLAIR and T1-CE in an intra-dataset setting. 
Across all scenarios, pDDPM-IQA consistently achieves state-of-the-art (SOTA) performance, with statistical significance (p $<$ 0.05), underscoring its robustness and adaptability to diverse imaging modalities.

\noindent\textbf{Effectiveness of $L_\text{FQ}$ and AIR-based Transformation.} Fig.~\ref{ablation_fig} (b) presents an ablation study on our two key components.
Introducing $L_\text{FQ}$ improves performance over the baseline, while AIR-based transformation further boosts results.
These highlight the effectiveness of $L_\text{FQ}$ and AIR-based transformation in enhancing anomaly detection.

\noindent\textbf{Generalization.}
To verify the generalization of our IQA approach, we apply it to another baseline, DDPM, and term it DDPM-IQA. We evaluate it on MSLUB T1 and T2, BraTS T1, T2, FLAIR, and T1-CE, using the same experimental settings as in Table~\ref{results_t2} and Fig.~\ref{ablation_fig} (a). As shown in Fig.~\ref{ablation_fig} (c), the IQA approach consistently enhances DDPM's performance across all datasets and modalities. These findings confirm that our IQA approach is broadly applicable and effective across various reconstruction-based anomaly detection methods.

\label{alpha_albation}
\noindent\textbf{$\alpha$ Sensitivity Study.} 
Fig.~\ref{ablation_fig} (d) shows the impact of $\alpha$ in Fusion Quality Loss. Instead of fine-tuning for each setting, we intentionally use a suboptimal yet effective $\alpha$. Even with $\alpha = 0.84$, our method consistently outperforms all baselines, showing its robustness and low sensitivity to $\alpha$ variations.

\section{Discussion and Conclusion}
In this study, we investigated reconstruction-based anomaly detection in brain MRI from an image quality assessment (IQA) perspective and proposed a novel IQA approach for medical anomaly detection. Our approach has two key components:
(1) a \textit{fusion quality loss} that combines SSIM with $\ell_1$ loss to better capture discrepancies between reconstructed and original images; and (2) an \textit{average intensity ratio (AIR)-based transformation} to amplify differences between normal and abnormal regions. 
Applied to a baseline pDDPM model (denoted pDDPM-IQA), our approach significantly outperforms state-of-the-art methods across multiple datasets and modalities.
It is worth noting that the proposed fusion quality loss and AIR-based data transformation are specific implementations under the broader IQA approach.
Therefore, further research into new metrics that better capture image anomalies than the current fusion quality loss could be a valuable direction.

\section*{Acknowledgment}

J. Chen was partially supported by NFDI4Bioimage, funded by the German Research Foundation (DFG) within the framework of the NFDI-project number 501864659, and also supported by the ``Ministerium für Kultur und Wissenschaft des Landes Nordrhein-Westfalen'' and ``Der Regierende Bürgermeister von Berlin, Senatskanzlei Wissenschaft und Forschung''.


\bibliographystyle{ieeetr}
\bibliography{Conference-LaTeX-template_10-17-19/BIBM,Conference-LaTeX-template_10-17-19/IEEEabrv}

\end{document}